\documentclass[twocolumn,aps,amsmath,amssymb,superscriptaddress,prl]{revtex4-1}   
\usepackage[latin2]{inputenc}
\usepackage{amsmath,amssymb}
\usepackage{amsfonts}
\usepackage{bm}
\usepackage{dcolumn}
\usepackage{setspace}
\usepackage{graphicx}
\usepackage{multirow}
\usepackage{verbatim}
\usepackage{epstopdf}
\usepackage[normalem]{ulem}

\def\bar{\begin{array}}
\def\ear{\end{array}}

\def\u{\uparrow}
\def\d{\downarrow}

\def\s{\sigma}
\def\f{\frac}

\def\nn{\nonumber}

\def\r{\mathbf{r}}

\def\k{\mathbf{k}}

\begin{document}

\title{Fock space embedding theory for strongly correlated topological phases}

\author{Ryan Requist} 
\affiliation{Max Planck Institute of Microstructure Physics, Weinberg 2, 06120, Halle, Germany}
\affiliation{
Fritz Haber Center for Molecular Dynamics, Institute of Chemistry, The Hebrew University of Jerusalem, Jerusalem 91904 Israel
}
\author{E. K. U. Gross}
\affiliation{Max Planck Institute of Microstructure Physics, Weinberg 2, 06120, Halle, Germany}
\affiliation{
Fritz Haber Center for Molecular Dynamics, Institute of Chemistry, The Hebrew University of Jerusalem, Jerusalem 91904 Israel
}

\date{\today}

\begin{abstract}
A many-body wave function can be factorized in Fock space into a marginal amplitude describing a set of strongly correlated orbitals and a conditional amplitude for the remaining weakly correlated part.  The marginal amplitude is the solution of a Schr\"odinger equation with an effective Hamiltonian that can be viewed as embedding the marginal wave function in the environment of weakly correlated electrons.  Here, the complementary equation for the conditional amplitude is replaced by a generalized Kohn-Sham equation, for which an orbital-dependent functional approximation is shown to reproduce the topological phase diagram of a multiband Hubbard model as a function of crystal field and Hubbard parameters.  The roles of band filling and interband fluctuations are elucidated.
\end{abstract}

\maketitle

First-principles calculations of topological invariants usually rely on the Kohn-Sham band structure.  This is problematic for correlated materials: the topological phase inferred from a mean-field band structure need not coincide with the actual topological phase determined from the correlated many-body wave function.  Although one can argue that a topological invariant cannot change as interactions are turned on adiabatically while maintaining an energy gap and the relevant symmetries, true strongly correlated topological phases could be characterized as precisely those phases that are not adiabatically connected to the noninteracting Kohn-Sham ground state.  

Embedding theories have been successful in describing electronic correlations beyond standard density functionals in extended systems \cite{anisimov1997, lichtenstein1998, govind1999, biermann2003, kotliar2006, held2006, koch2008, knizia2012, welborn2016, senjean2018, fertitta2018, rusakov2019, senjean2019}.  In most embedding schemes, a real-space fragment such as an impurity site or a small set of atomic or molecular orbitals, is embedded in its surroundings.  The use of a real-space fragment, typically with only local or short-range interactions, inherently limits the nonlocality and hence quasimomentum dependence that can be described.  Since topological invariants depend on the global $\k$-dependence of the state, either through the twisting of Bloch functions in the Brillouin zone \cite{thouless1982} or the behavior of the many-body wave function under twisted boundary conditions \cite{laughlin1981,niu1984,niu1985}, it is natural to ask whether alternative embedding theories might be better suited to capturing momentum-dependent correlations.

This Letter proposes a novel embedding theory rooted in the exact factorization (EF) methodology  \cite{hunter1975,gidopoulos2014,abedi2010}, a scheme for splitting the many-body wave function into marginal and conditional probability amplitudes describing different degrees of freedom. 
It has been applied to electrons and nuclei \cite{abedi2010, gidopoulos2014, min2014, min2015, requist2016a}, fast and slow electrons \cite{schild2017}, electrons and photons \cite{hoffmann2018,lacombe2019b}, and electrons and phonons \cite{requist2019b}.  We turn to the problem of strong electron-electron correlation and use an extension of the exact factorization formalism to Fock space \cite{gonze2018} to develop a novel embedding theory.  Here, the marginal amplitude describes the strongly correlated degrees of freedom embedded in the remaining weakly correlated degrees of freedom.

For the purpose of calculating topological invariants, the key advantage of an EF-based embedding formalism lies in the ability of the explicitly correlated marginal wave function to capture the $\k$-dependent phase information of the strongly correlated electrons, which is partly lost in approaches based on Green's functions or reduced density matrices.  The contribution of the remaining weakly correlated electron bands can be adequately described through mean-field Bloch functions.  Thus, one can go beyond density functional theory, while avoiding empirical models and the infeasibility of including all degrees of freedom in a many-body calculation of the solid. 

To apply the exact factorization formalism to a many-electron wave function, we start by writing it in Fock space as a superposition of products of many-body configurations
\begin{align}
|\Psi\rangle &= \sum_{SD} c_{SD} |S\rangle |D\rangle {,} \label{eq:superposition}
\end{align}
where $|S\rangle = c_{s_1}^{\dag} \ldots c_{s_{N_S}}^{\dag}|0\rangle$ and $|D\rangle = c_{d_1}^{\dag} \ldots c_{d_{N_D}}^{\dag}|0\rangle$ are constructed from orbitals belonging to mutually orthogonal sets $\mathcal{S}$ and $\mathcal{D}$ of weakly  and strongly correlated orbitals; $S=s_1 s_2\ldots s_{N_S}$ is a string of indices labeling the orbitals and similarly for $D$.  The $|S\rangle$ and $|D\rangle$ factors may have varying particle numbers in each term of Eq.~(\ref{eq:superposition}), but $N_S+N_D=N$ is fixed.  Following the EF procedure \cite{gonze2018}, the marginal amplitude is defined to be $\chi_D=e^{i\Theta_D}|\chi_D|$ with
\begin{align}
|\chi_D|^2 = \sum_S |c_{SD}|^2
\end{align}
and an arbitrary phase $\Theta_D$.  The conditional factor
\begin{align}
\Phi_{S|D} = c_{SD}/\chi_D 
\end{align}
then satisfies the partial normalization condition
\begin{align}
\sum_S |\Phi_{S|D}|^2 = 1 \; \forall \; D {.}
\end{align}
We thus arrive at the factorization $c_{SD} = \chi_D \Phi_{S|D}$. 

A criterion is needed to partition the complete set of single-particle orbitals into weakly and strongly correlated sets $\mathcal{S}$ and $\mathcal{D}$.  While different strategies are possible, here we perform the separation through a criterion involving the natural occupation number bands, i.e.~the bands formed by the $\k$-dependent eigenvalues of the one-body reduced density matrix in the Brillouin zone of the crystal.  Strongly correlated orbitals are defined to be those belonging to a band whose occupation numbers satisfy $f_{lower}\leq f_{n\k}\leq f_{upper}$ with judiciously chosen $f_{lower}$ and $f_{upper}$, while the rest are called weakly correlated.  While our formalism can be applied with any choice of $\mathcal{S}$ and $\mathcal{D}$, it will become computationally prohibitive if too many bands are included in $\mathcal{D}$.  We have in mind situations where correlations are concentrated in relatively few bands with only weak residual correlations in $\mathcal{S}$, so that the occupation numbers of the latter are very close to 0 and 1 and can be accurately described by standard density functional approximations. Natural occupation numbers in extended systems have only been reported for one-band systems, namely the homogeneous electron gas \cite{ortiz1994a, ortiz1997, lathiotakis2007} and the one-dimensional Hubbard model \cite{koch2001,rusakov2016}.  Recent calculations of a multiband Hubbard model \cite{requist2019a} used an unfolding procedure \cite{requist2018} with twisted boundary conditions to derive a continuous band structure from the discrete set of natural occupation numbers and orbitals obtained from exact diagonalization.  It was found that when there is a disparity in the strength of interactions in bands of different orbital character one can have simultaneously a set of strongly correlated bands satisfying $f_{lower}\leq f_{n\k}\leq f_{upper}$ and another set with occupation numbers very close to 0 and 1. This situation might arise, for instance, in transition metal-bearing oxides, where bands with predominantly transition metal $d$ orbital character experience a stronger Hubbard repulsion.  In general, it might be necessary carry out multiple self-consistent calculations with different partitions to find the variational minimum.

\begin{figure}[htb!]
\centering
\includegraphics[width=0.90\columnwidth]{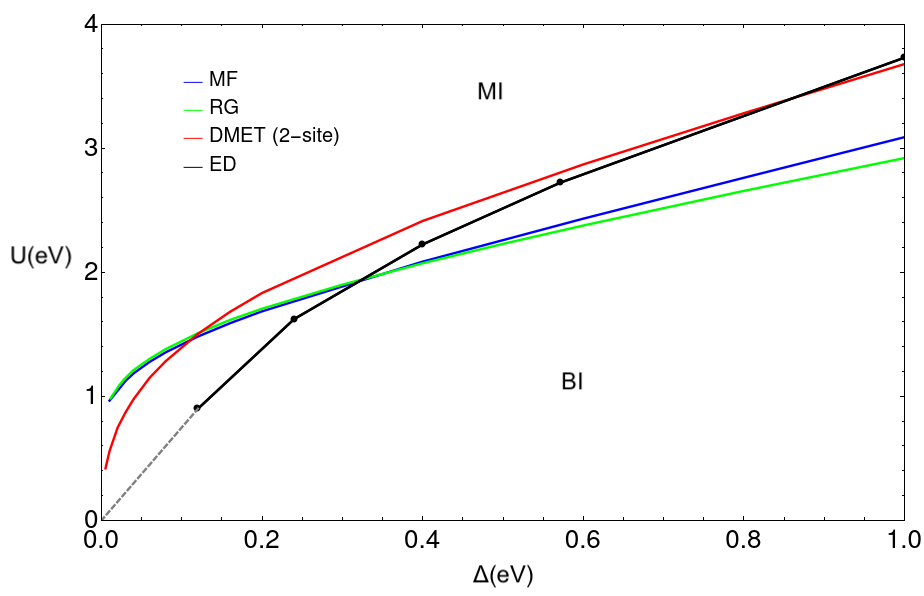} 
\caption{Phase diagram of the half-filled ionic Hubbard model.  The phase boundary between band insulator (BI) and Mott insulator (MI) phases from mean-field theory (MF, blue), bosonization+renormalization group (RG, green) from Ref.~\onlinecite{tsuchiizu1999}, density matrix embedding theory (DMET, red), and exact diagonalization (ED, black); dashed line extrapolates to the homogeneous model; hopping $t=1$~eV.}
\label{fig:dmet} 
\end{figure}

Given the above choice of partition, our theory embeds a set of natural Bloch orbital bands in an environment made up of all remaining bands.  This is the crux of our approach and distinguishes it from all other embedding theories, most of which rely on a real-space partition.  Preserving translational symmetry by keeping entire bands intact in the correlated subspace is the key to reliably calculating topological invariants and, in turn, topological phase diagrams, which depend on nonlocal correlations beyond those confined within a real-space fragment.  To see the effect of nonlocal correlations, in Fig.~\ref{fig:dmet} we compare the phase diagram of the half-filled ionic Hubbard model $\hat{H}=\sum_{i\s} [-t c_{i\s}^{\dag} c_{i+1\s} + H.c. + (-1)^i \Delta c_{i\s}^{\dag} c_{i\s}] + \sum_i U \hat{n}_{i\u} \hat{n}_{i\d}$ calculated by several methods: mean-field theory (MF),  a renormalization group (RG) method applied to the bosonized Hamiltonian \cite{tsuchiizu1999}, density matrix embedding theory (DMET) \cite{knizia2012}, and exact diagonalization (ED) extrapolated to the thermodynamic limit using data from periodic 8-, 10- and 12-site models following the approach in Ref.~\onlinecite{gidopoulos2000}.  In our DMET implementation with a 2-site embedding fragment and spin-symmetry preserving interacting bath \cite{wouters2016}, the phase boundary calculated from the polarization of the band electrons is overestimated, i.e.~the band insulator to Mott insulator transition occurs at a higher $U$ than in the exact result.  This demonstrates that nonlocal correlations can be important even when a model contains only local (Hubbard) interactions. 

A Schr\"odinger-like equation for the marginal factor is derived by inserting $c_{S'D'} = \chi_{D'} \Phi_{S'|D'}$ into the original Schr\"odinger equation with Hamiltonian $\hat{H}$, multiplying on the left by $\Phi_{S|D}^*$ and summing over $S$ and $S'$ to obtain
\begin{align}
\sum_{D'} H_{DD'} \chi_{D'} = E \chi_D {,} \label{eq:marginal}
\end{align}
where 
\begin{align}
H_{DD'} = \sum_{SS'} \Phi_{S|D}^* H_{SD|S'D'} \Phi_{S'|D'}  \label{eq:H:marginal}
\end{align}
will be referred to as the embedding Hamiltonian.  The factor $H_{SD,S'D'} = \langle SD | \hat{H} | S'D'\rangle$, which can be broken down into one-body and two-body contributions, induces charge fluctuations between $\mathcal{S}$ and $\mathcal{D}$ subspaces.  Despite its simple appearance, the Hamiltonian in Eq.~(\ref{eq:H:marginal}) is actually quite unusual in that it couples many-body configurations with vastly different particle number, spin, etc.  

The next step would be to derive the equation for the conditional factor $\Phi_{S|D}$, which is needed to explicitly construct $H_{DD'}$.  However, solving the coupled equations for $\chi$ and $\Phi_{S|D}$ in their full complexity would be tantamount to solving the original Schr\"odinger equation.  Thus, we seek an alternative path that will determine $H_{DD'}$ as well as the strongly  and weakly correlated orbitals self-consistently.  To obtain a scheme that can be applied to real materials, we couple Eq.~(\ref{eq:marginal}) to the following generalized Kohn-Sham (GKS) equation:
\begin{align}
\left[ \f{\hat{p}^2}{2m} + \hat{v}_{ext} + \hat{v}_{hxc} + \hat{w}_{hxc} \right] |\psi_{n\k}\rangle = \epsilon_{n\k} |\psi_{n\k}\rangle {,} \label{eq:KS}
\end{align}
where $\hat{v}_{hxc} = \hat{v}_{hxc}[n,\psi_{d\k},\chi]$ denotes a scalar multiplicative potential and $\hat{w}_{hxc}[n,\psi_{d\k},\chi]$ is a nonlocal operator acting only in the subspace of strongly correlated natural orbitals $\psi_{d\k}\in \mathcal{D}$.  Both $\hat{v}_{hxc}$ and $\hat{w}_{hxc}$ are functionals of the electronic density $n(\r)$, $\psi_{d\k}(\r)$, and  $\chi_D$.  The Hamiltonian matrix elements $H_{DD'}$ are similarly functionals of $n(\r)$ by virtue of the Hohenberg-Kohn theorem \cite{hohenberg1964}.  The GKS equation can be derived by making the energy stationary with respect to variations of $n(\r)$ and $\psi_{d\k}$ \cite{requist2019a}.

The density is determined in a nonstandard way as
\begin{align}
n(\r) = \sum_{n\k\s\in \mathcal{S}} f_{n\k} |\psi_{n\k\s}(\r)|^2 + \sum_{d\k\s\in \mathcal{D}} f_{d\k} |\psi_{d\k\s}(\r)|^2 \label{eq:n}
\end{align}
with fractional strongly correlated occupation numbers determined from the marginal factor according to $f_{d\k} = \langle \chi | c_{d\k\s}^{\dag} c_{d\k\s} | \chi\rangle$.  The weakly correlated orbitals $\psi_{n\k}\in\mathcal{S}$ have occupation numbers 0 and 1, with the possible exception of orbitals with energies equal to the chemical potential.  The coupling to the marginal equation enters implicitly through $n(\r)$ as well as the $\chi$-dependence of $v_{hxc}(\r)=\left.\delta E_{hxc}/\delta n(\r)\right|_{\phi_{d\k}}$ and the matrix elements 
\begin{align}
\langle \psi_{d\k} | \hat{w}_{hxc} | \psi_{d'\k} \rangle = \f{\left.\Big< \psi_{d'\k'} \Big| \f{\delta E_{hxc}}{\delta\psi_{d\k}^*} \Big>\right|_{n} - \left.\Big< \f{\delta E_{hxc}}{\delta\psi_{d'\k'}} \Big| \psi_{d\k} \Big>\right|_{n}}{f_{d\k}-f_{d'\k'}} {,}
\end{align}
where the energy has been partitioned as $E = T_{s,e} + \int v(\r) n(\r) d\r + E_{hxc}$ with an ensemble kinetic energy functional $T_{s,e}$ defined through the constrained search \cite{levy1979} over ensembles of Slater determinants $\rho_s$.  

So far no approximations have been made.  Solving Eqs.~(\ref{eq:marginal}) and (\ref{eq:KS}) self-consistently would yield the exact $n(\r)$, $\psi_{d\k}(\r)$, and  $\chi_D$.  To have a practical scheme, we need to specify functional approximations for $v_{hxc}(\r)$, $\hat{w}_{hxc}$ and $H_{DD'}$.  For this purpose, we introduce the following approximation, which we call the Aufbau approximation, to define the conditional amplitude $\Phi_{S|D}$.  Namely, for each configuration $|D\rangle$ we define the Slater determinant $|S^{\rm Aufbau}\rangle$ built from the $N_S=N-N_D$ lowest energy weakly correlated orbitals subject to the conditions that (i) the $\hat{S}_z$ eigenvalues satisfy $M_S+M_D=M$ and (ii) the quasimomentum eigenvalues satisfy $\mathbf{K}_{D}+\mathbf{K}_{S}=\mathbf{K}$, where $M$ and $\mathbf{K}$ are the quantum numbers of $|\Psi\rangle$ (additional symmetries could also be imposed at this stage).  Since the multi-index $D$ uniquely determines $S$ if the weakly correlated orbitals are nondegenerate, which, for simplicity, we assume, there exists a function $S^{\rm Aufbau}(D)$.  Thus, we define
\begin{align}
\Phi_{S|D} = \left\{ \bar{ll} 1 & \textrm{if}\;\;S=S^{\rm Aufbau}(D) \\
0 & \mathrm{otherwise} \ear \right. {.} \label{eq:Phi}
\end{align} 
Since $|S^{\rm Aufbau}\rangle$ is a Slater determinant of KS-like orbitals, Eq.~(\ref{eq:Phi}) allows us to construct $H_{DD'}$ as well as the total energy as implicit functionals of $n(\r)$.  Thus, for any given choice of approximate local potential $\hat{v}_{hxc}$, we have specified an approximation that can be applied to real materials without further functional development.   

Although solving the many-body Schr\"odinger equation in Eq.~(\ref{eq:marginal}) remains a challenging task, especially in higher dimensions, it is worth emphasizing that the EF method has simplified the original problem to a degree that established many-body techniques can be applied, while retaining the coupling to all remaining electronic degrees of freedom of the solid.  

Before pursuing calculations of topological phases in real materials, it is desirable to test the theory in a case where it can be compared with benchmark calculations. To this end, we calculate the topological phase diagram of a multiband ionic Hubbard model, comprising two $s$ bands and two $d$ bands:
\begin{align}
\hat{H}_s &= -\sum_{i\s} (t_{s,ii+1}(\xi) c_{i\s}^{\dag} c_{i+1\s} + H.c.) + \sum_{i\s} \epsilon_{s,i} c_{i\s}^{\dag} c_{i\s} {,} \nn \\
\hat{H}_d &= -\sum_{i\s} (t_{d,ii+1}(\xi) d_{i\s}^{\dag} d_{i+1\s} + H.c.) + \sum_{i\s} \epsilon_{d,i} d_{i\s}^{\dag} d_{i\s} \nn \\
& \quad + U \sum_i \hat{n}_{d,i\u} \hat{n}_{d,i\d} {.}  \label{eq:H:1}
\end{align}
To study the behavior of bands with vastly disparate interactions, we take the $s$ electrons to be noninteracting.  The hopping amplitudes depend on the sublattice displacement $\xi$ according to $t_{s,ii+1}=t_{s1}=t_0-2g_s \xi$ for $i=$ odd and $t_{s,ii+1}=t_{s2}=t_0+2g_s \xi$ for $i=$ even and similarly for $t_{d,ii+1}$.  The onsite potentials are staggered, i.e.~$\epsilon_{s,i}=(-1)^i \Delta_s$ and similarly for $\epsilon_{d,i}$.  The $s$ and $d$ bands are coupled by a hopping term
\begin{align}
\hat{H}_{sd} &= -t_{sd} \sum_{i\s} (c_{i\s}^{\dag} d_{i+1\s} + d_{i\s}^{\dag} c_{i+1\s} + H.c.)  {.} \label{eq:H:parts}
\end{align}
This model has a band insulator to Mott insulator transition at a critical value of the Hubbard parameter $U$, similar to the single band model \cite{nagaosa1986a,resta1995,schoenhammer1995,ortiz1996,fabrizio1999,aligia1999,gidopoulos2000,torio2001,wilkens2001,takada2001,lou2003,kampf2003,manmana2004,tsuchiizu2004,tincani2009}.  

We also include a crystal field term $\hat{\Delta}_{sd} = \Delta_{sd} \sum_{i\s} (n_{d,i\s} -  n_{s,i\s})$ to break particle-hole symmetry.  By varying $\Delta_{sd}$, we control the filling of the $d$ band and study the effect of band-filling on the quantum phase transition.  We are effectively using the weakly correlated ``spectator" band to dope the strongly correlated band (reminiscent of carrier doping in cuprate superconductors \cite{imada1998}).  
Previous studies involving variable band-filling in multiband Hubbard models, e.g.~in connection with the orbital-selective Mott transition \cite{hotta2002,anisimov2002,fang2004,koga2004,werner2007,demedici2009} and strongly correlated superconductivity \cite{capone2004}, have tended to treat higher symmetry scenarios with the same value of intraband $U$ for all bands and additional interband and exchange interactions. 

\begin{figure}[tb!]
\centering
\includegraphics[width=0.86\columnwidth]{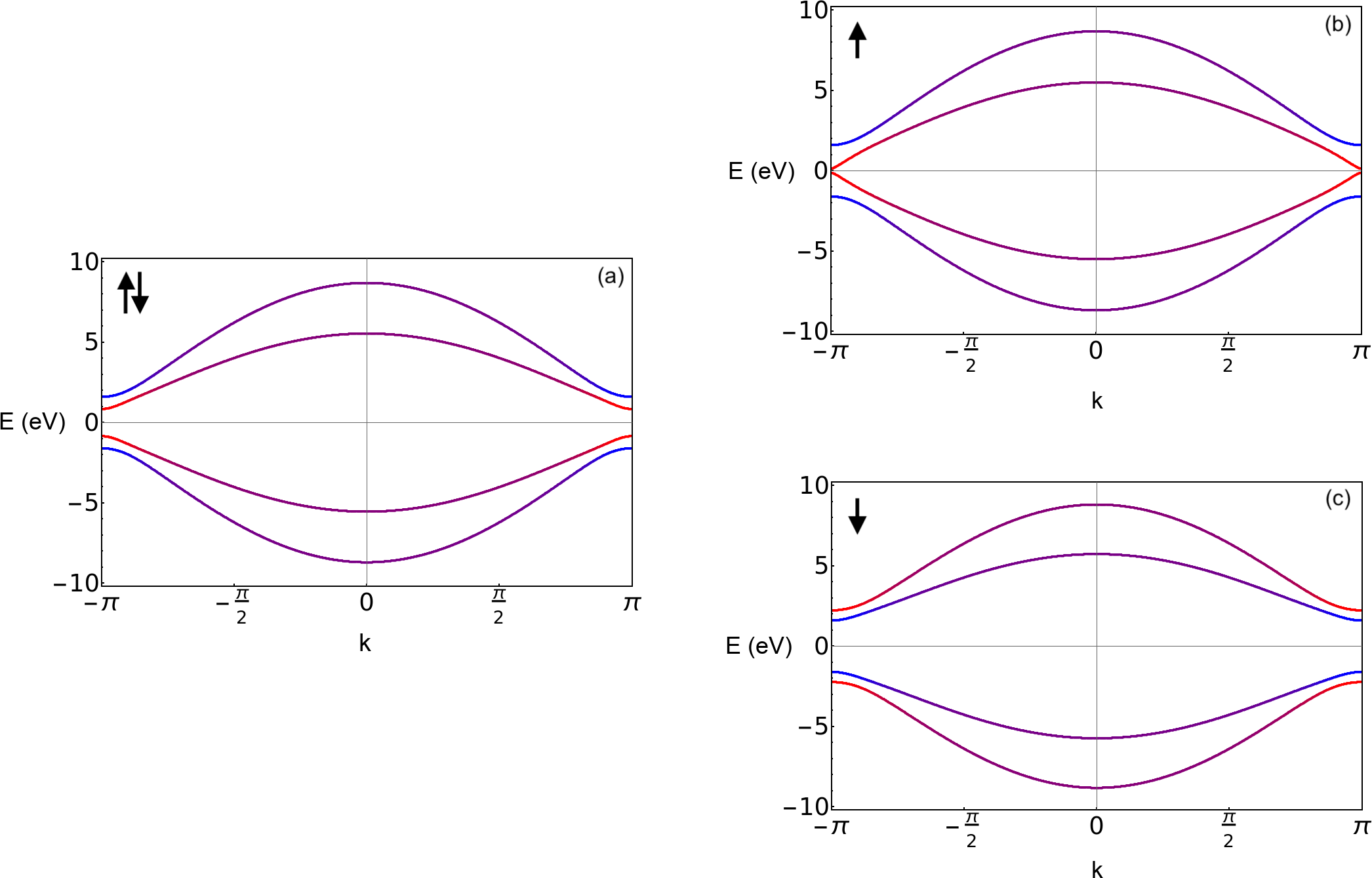} \\
\includegraphics[width=0.86\columnwidth]{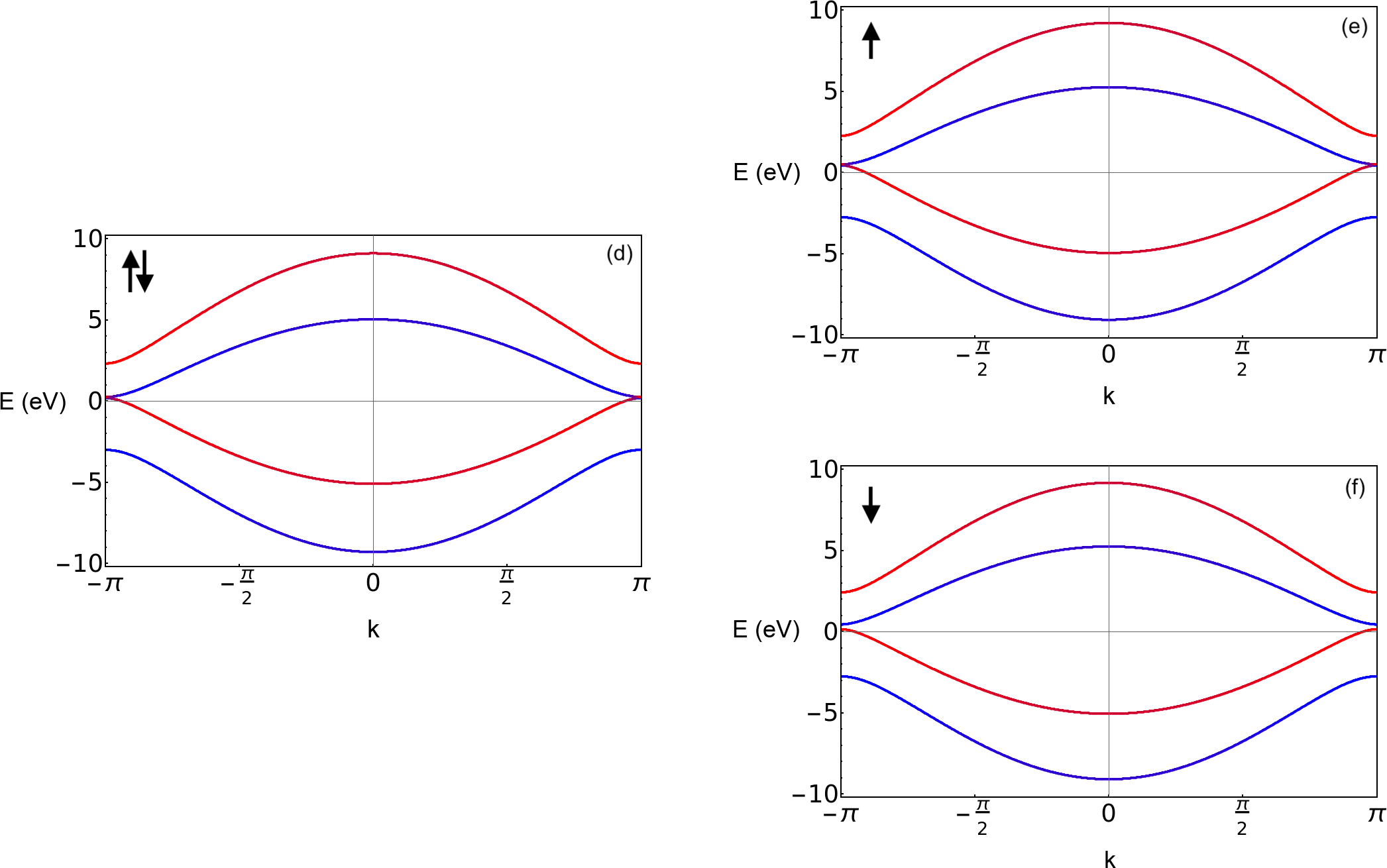} 
\caption{(a-c) Interaction-driven symmetry breaking transition in mean-field band structures for $U = 8.28$~eV and $\Delta_{sd}=0$. (d-f) Crystal field-driven transition for $U=6.6$~eV and $\Delta_{sd}=1.4$~eV.  In both cases, $t_s=t_d=3.5$, $\Delta_s=1.6$, $\Delta_d=2.0$, $\xi=5 \times 10^{-5}$ and $t_{sd} =  0.8$; all in eV.  Color scale (blue to red) indicates the orbital character ($s$ to $d$).}
\label{fig:mean-field:U} 
\end{figure}

We start by discussing the model from the mean-field perspective.  The interaction-driven transition from paramagnetic to antiferromagnetic phase upon increasing $U$ is depicted in Fig.~\ref{fig:mean-field:U}a-c.  The closing of the $d$-orbital up-spin gap (Fig.~1b) as the critical value of $U$ is approached from above signals the transition to the paramagnetic phase.  A different scenario is found for the crystal field-driven transition (Fig.~\ref{fig:mean-field:U}d-f) with varying $\Delta_{sd}$, where the gap-closing defining the critical point occurs between bands of different orbital character (see Fig.~1e).

\begin{figure}[tb!]
\centering
\includegraphics[width=0.85\columnwidth]{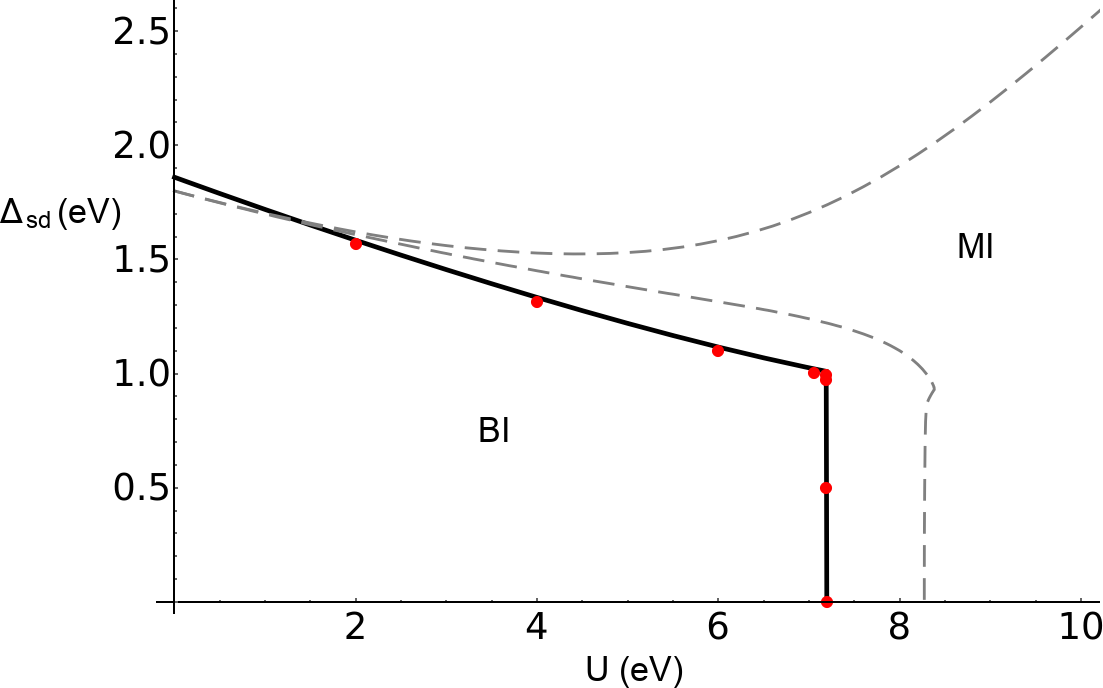} 
\caption{Phase diagram of the two-orbital Hubbard model as a function of the Hubbard and crystal field parameters $U$ and $\Delta_{sd}$.  Boundaries between band insulator (BI) and Mott insulator (MI) phases are calculated with the Aufbau approximation (solid black curve), exact diagonalization (red dots) and the mean-field approximation (dashed gray curves) for $t_s=t_d=3.5$, $\Delta_s=1.6$, $\Delta_d=2.0$, $\xi=5\times 10^{-5}$ and $t_{sd} =  0.8$  (all in eV).}
\label{fig:phase:diagram} 
\end{figure}

Figure \ref{fig:phase:diagram} shows the topological phase diagram obtained in the Aufbau approximation as a function of $U$ and $\Delta_{sd}$.  The Aufbau solution (solid black curve) displays transitions from a band insulator (BI) to a Mott insulator (MI) as either $U$ or $\Delta_{sd}$ is increased.  The phase boundaries are determined from jumps of $\pi$ in the marginal geometric phase $\gamma_{\chi} = \int_{0}^{2\pi} i \langle \chi | \partial_{\alpha} \chi \rangle d\alpha$ \cite{suppl}.  The phase transitions signal a discontinuous change in the macroscopic polarization $P=-(e/2\pi)\gamma$, which is a topological invariant quantized to $P=0$ or $\f{e}{2}\mod e$ by the parity symmetry of the model.  Similar behavior is well known in single-orbital ionic Hubbard models in one and two dimensions \cite{resta1995,ortiz1996,gidopoulos2000}.  
We use a small symmetry breaking dimerization $\xi=5\times 10^{-5}$, and the Born-von K\'arm\'an cells used in our calculations are too small to see an intermediate bond-ordered phase \cite{fabrizio1999,wilkens2001,tsuchiizu2004}.
The Aufbau solution agrees well with the one obtained by numerical exact diagonalization (red dots), demonstrating the viability of the approximation.

The BI-MI transition is reflected in the paramagnetic to antiferromagnetic mean-field phase boundaries (dashed gray curves), which roughly follow the BI to MI transition of the correlated solution.  However, a second symmetry-restoring transition is reached when $\Delta_{sd}$ is further increased.  For $U=0$, the crystal field-driven transition occurs at exactly $\Delta_{sd}=\f{1}{2}(\Delta_s+\Delta_d)=1.80$~eV, as correctly reproduced in the mean-field solution.  The intercept deviates slightly in the Aufbau approximation, which does not become exact in the limit $U\rightarrow 0$ because it does not capture all $t_{sd}$-induced interband charge fluctuations.  At the symmetry-breaking transition, the geometric phase of the $d$-orbital valence band associated with one spin, e.g.~the down-spin, $\gamma_{d\d} = \int_{-\pi/a}^{\pi/a} i \langle u_{dk\d} |\partial_k u_{dk\d}\rangle dk$, jumps from $\pi$ to $0$ as $\Delta_{sd}$ (or $U$) is increased.  At the symmetry-restoring transition, the geometric phase of the opposite spin, $\gamma_{d\u}$, also jumps from $\pi$ to $0$.  No such second transition was observed in either the exact diagonalization or Aufbau solutions in the investigated range. 

\begin{figure}[htb!]
\centering
\includegraphics[width=0.48\columnwidth]{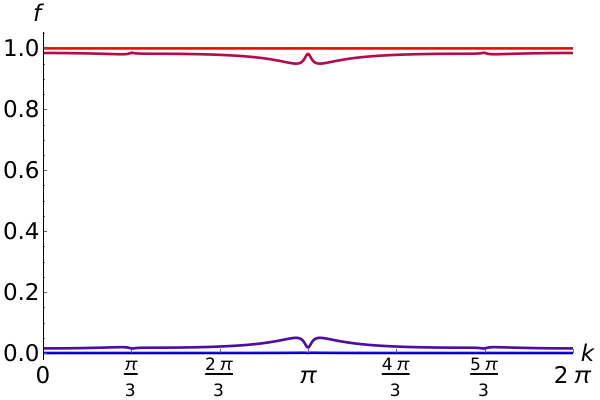} \includegraphics[width=0.48\columnwidth]{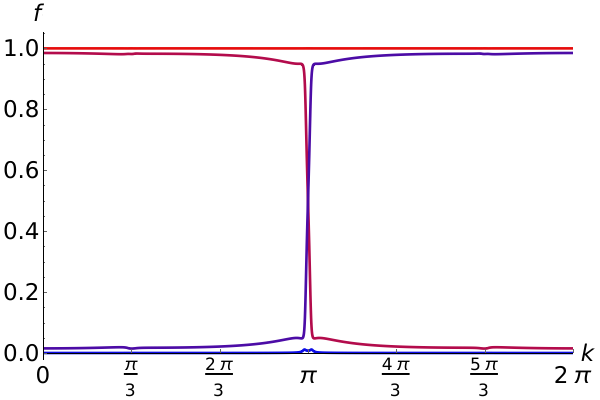} 
\caption{Inversion of natural occupation number bands as $\Delta_{sd}$ is increased through the phase boundary from $\Delta_{sd}=1.0$ to $\Delta_{sd}=1.1$ for $t_s=t_d=3.5$, $\Delta_s=1.6$, $\Delta_d=2.0$, $\xi=5\times 10^{-5}$, $t_{sd} =  0.8$ and $U=6.6$ (all in eV). The zone boundary ($k=\pi$) is positioned at the center to make the crossing visible.}
\label{fig:occupation:bands} 
\end{figure}

The change in the topological invariant from the BI to MI phase coincides with a change in the topology of the natural occupation number bands as shown in Figure \ref{fig:occupation:bands}.  The strongly correlated occupation number bands develop a crossing at the zone boundary, which implies a natural occupation number ``band inversion."   We have described this phenomenon in the Rice-Mele-Hubbard \cite{requist2017}, and it is similar to the purity-gap closing studied in the quench dynamics of ultracold atoms \cite{kruckenhauser2018}.  With respect to other descriptors of the transition, the natural occupation number bands have the virtues that they build in the crystal symmetries through their symmetry properties and those of the natural Bloch orbitals and can be straightforwardly extended to probe real-time dynamics.  

The solution of the embedding Schr\"odinger equation, where different marginal charge states are combined into a single state vector $|\chi\rangle$, is fundamentally different from the solution of an effective $d$-electron Hamiltonian with band-filling controlled by a chemical potential.  While in the former case the phase boundaries can be detected by discontinuous $\pi$-jumps, the mean-field geometric phase of the latter is blind to the paramagnetic-antiferromagnetic transition \cite{suppl}.  This underscores the usefulness of an EF approach built on pure states for the detection of topological phase transitions.  We expect our approach to also yield accurate energies and local observables.

In summary, a novel embedding theory based on a Fock space factorization has been found to  reproduce the topological phase diagram of a strongly correlated multiband system.  Calculations employing the proposed Aufbau approximation can be expected to aid the ongoing search for novel strongly correlated materials.

{\it Note added.} We recently became aware of related work by Lacombe and Maitra that also develops an exact factorization-based embedding method \cite{lacombe2020}.

\bibliography{bibliography2019aa}

\end{document}


{\Large \textbf{Supplemental material}}

\section{Many-body calculations}

Both the Aufbau and exact diagonalization solutions reported in Fig.~2 of the main text were carried out for a Born-von K\'arm\'an cell with length $L=3a$.  With 4 orbitals (2 atoms and 2 orbitals/atom) in each primitive cell, this corresponds to 12 electrons and 12 orbitals.  Brillouin zone integrations and twist-averaging were performed with between 96 and 384 $k$ points.  

The solution in the Aufbau approximation was obtained by minimizing the energy with respect to $\chi$ and all orbitals (weakly and strongly correlated).  Since the $s$ band in our model is noninteracting (see Eq.~(13) of the main text), we can avoid solving the GKS equation.  This minimization procedure results in fractional occupation numbers for the weakly-correlated orbitals, which can therefore no longer be identified with a noninteracting KS system with a local potential.  Nevertheless, since their occupation numbers are all very close to 0 or 1, namely within $1.5\times 10^{-5}$ for $t_s=3.5$, $t_d=3.0$, $\Delta_s=0.50$, $\Delta_d=0.55$, $\Delta_{sd}=0$, $\xi=0.0080$, $t_{sd}=0.8$ and $U=7.2$ (all in eV), the result is essentially identical to the result of solving the GKS equations with integer-occupied weakly-correlated orbitals. 

The exact diagonalization solution was obtained by diagonalizing the Hamiltonian in a restricted Hilbert space consisting of all states of the following types: 
\begin{enumerate}[(a)]
\item $|FS\rangle \otimes |D\rangle$ with $N_D=6$, $K_D=0$, $M_D=0$ 
\item $a_{vk\s}|FS\rangle \otimes |E\rangle$ with $N_E=7$, $K_E+K_S=0$, $M_E+M_S=0$
\item $a_{uk\s}^{\dag}|FS\rangle \otimes |F\rangle$ with $N_F=5$, $K_F+K_S=0$, $M_F+M_S=0$
\item $a_{uk\s}^{\dag} a_{vk'\s}|FS\rangle \otimes |I\rangle$ with $N_I=6$, $K_I+K_S=0$, $M_I+M_S=0$
\item $a_{uk\s}^{\dag} a_{vk'\overline{\s}}|FS\rangle \otimes |J\rangle$ with $N_J=6$, $K_J+K_S=0$, $M_J+M_S=0$ {,}
\end{enumerate} 
where $|FS\rangle$ is the $s$ electron Fermi sea, $N_D$ is the number of electrons in configuration $|D\rangle$, $K_D$ is the quasimomentum eigenvalue, and $M_D$ is the $\hat{S}_z$ eigenvalue.

\section{Mean-field phase diagram of multiband vs.~single band models}

\begin{figure}[th!]
\includegraphics[width=0.95\columnwidth]{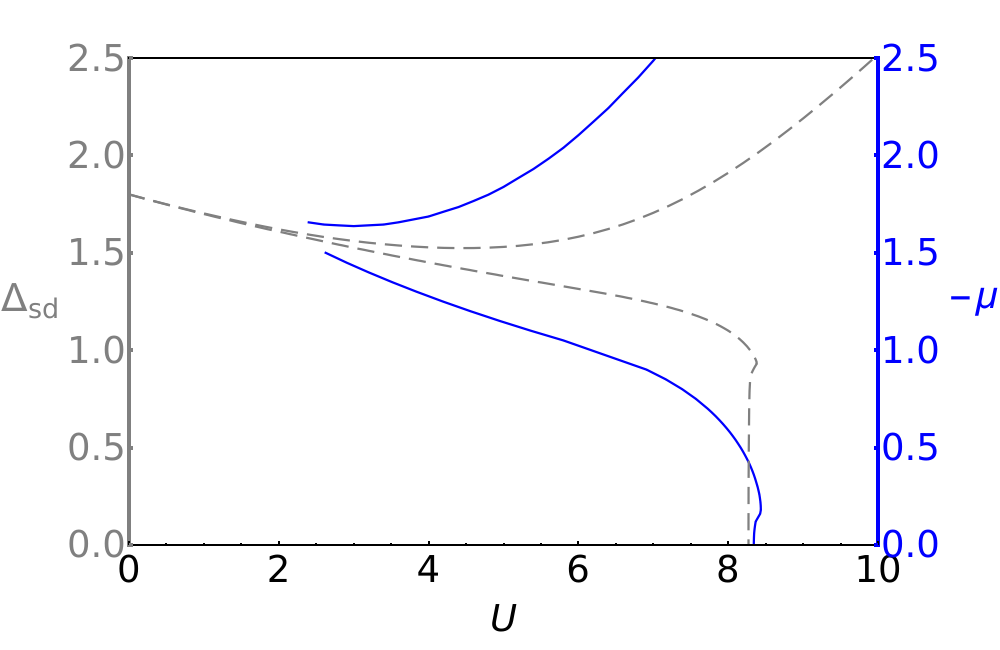} 
\caption{Mean-field phase diagram of the multiorbital Hubbard model of the main text (dashed gray curves) as a function of $(U,\Delta_{sd})$ and the corresponding single-orbital Hubbard model (blue curves) as a function of $(U,\mu)$. Parameters are $t_s=t_d=3.5$, $\Delta_s=1.6$, $\Delta_d=2.0$, $\xi=5 \times 10^{-5}$ and $t_{sd} =  0.8$; all in eV.}
\label{fig:mean-field}
\end{figure}

The crystal field parameter $\Delta_{sd}$ allows us to control the relative occupation of $s$ and $d$ orbitals.  The ground state, $|\chi\rangle$, of the embedding Hamiltonian in Eq.~(5) of the main text superposes different charge states of the strongly-correlated subspace into a single coherent pure state.  This is distinctly different from reduced density matrix methods, where the strongly-correlated subspace is described by a partially incoherent mixed state and its mean occupancy is controlled by a chemical potential.  To explore whether this difference has any consequences for the calculation of topological phase diagrams, in Fig.~\ref{fig:mean-field} we compare the mean-field phase diagrams of (i) the multiband Hamiltonian defined in Eqs.~(11) and (12) of the main text and (ii) the corresponding single-band Hamiltonian defined by $\hat{H}_d$ in Eq.~(11).   In case (i), the occupancy of the $d$ orbitals is controlled by the parameter $\Delta_{sd}$; in case (ii), it is controlled by the chemical potential $\mu$.  The significant differences in the phase diagrams of the two cases suggests that doping by a ``spectator" band is not equivalent to doping by an external particle reservoir.  Therefore, care should be taken when using the chemical potential as a parameter in a topological phase diagram.  There is another important distinction between the two cases that relates to their sensitivity to topological properties.  In case (i), the paramagnetic-antiferromagnetic phase boundaries is signaled by $\pi$-jumps in the up-spin and down-spin mean-field geometric phases, in case (ii) it is not.  In case (ii), the phase boundaries have instead been determined by the spin-symmetry-breaking paramagnetic to antiferromagnetic transition.

\section{Entanglement}

With the help of the Schmidt decomposition
\begin{align}
|\Psi\rangle &= \sum_{i=1}^{\mathcal{N}} \sqrt{\lambda_i} |\xi_i\rangle |\eta_i\rangle {,} 
\end{align}
the entanglement between a subsystem and its environment can be quantified by the entanglement entropy
\begin{align}
\mathbb{S} = -\sum_{i=1}^{\mathcal{N}} \sqrt{\lambda_i} \ln \sqrt{\lambda_i} {,}
\end{align}
where $\mathcal{N}$ is the number of nonzero eigenvalues $\lambda_i$ of the subsystem reduced density matrix
\begin{align}
\rho_{DD'} = \sum_S c_{SD} c_{SD'}^* {.}
\end{align}
The Aufbau approximation proposed in the main text delivers a wave function in the Schmidt-decomposed form 
\begin{align}
|\Psi^{\rm Aufbau} \rangle &= \sum_D \chi_D |D\rangle |S^{\rm Aufbau}(D)\rangle {.}
\end{align}
Every many-body configuration $|D\rangle$ is paired with a state $|S^{\rm Aufbau}(D)\rangle$ from the environment.  
The entanglement entropy in the Aufbau approximation is 
\begin{align}
\mathbb{S}^{\rm Aufbau} = -\sum_{i=1}^{\mathcal{N}_D} \chi_i \ln \chi_i {,}
\end{align}
where $\mathcal{N}_D$ is the number of nonzero coefficients $\chi_D$ and we have introduced an ordinal index, denoted $i$, which labels the set of all nonzero $\chi_D$ coefficients.  Since many of the $\chi_D$, defined by the solution of the model Hamiltonian in Eq.~(5) of the main text, differ significantly from 0 and 1 in the strongly correlated regime---a fact which follows from our assumption that the natural occupation numbers of the strongly correlated bands satisfy $f_{lower}\leq f_{n\mathbf{k}} \leq f_{upper}$---our embedding theory in the Aufbau approximation captures a significant amount of entanglement between the weakly and strongly correlated subspaces.

\section{Geometric phase}

Similar to many topological invariants \cite{niu1984}, the many-body macroscopic polarization can be evaluated in terms of the ground state of a twisted Hamiltonian, i.e.~with an artificial magnetic flux $\alpha$ \cite{ortiz1994b}.  The geometric phase of the Aufbau ground state is
\begin{align}
\gamma &= \sum_D \int_{-\pi}^{\pi} i \chi_D^* \partial_{\alpha} \chi_D d\alpha \nn \\
&\quad+ \sum_D |\chi_D|^2 \sum_{nk\s\in s(D)} \int_{-\pi/a}^{\pi/a} i \langle u_{nk\s} | \partial_k u_{nk\s} \rangle dk {,}
\end{align}
where $\alpha$ is an artificial magnetic flux and $u_{nk\s}(r)$ is the cell-periodic part of the Bloch function $\psi_{nk\s}(r)$.  The first term is simply the many-body geometric phase \cite{ortiz1994b} of the marginal factor, while the second term is a $|\chi_D|^2$-weighted sum of single-particle contributions \cite{king-smith1993}.  For clarity, the formula is given for the special case of a one-dimensional system. 

\bibliography{bibliography-suppl}